\newcommand{\edit}[1]{#1}

\documentclass[]{IEEEapm}

\jmonth{}
\pubyear{}

\begin{document}

\title{Fast approximate solvers for metamaterials design in electromagnetism.}

\author{R.~Pestourie}

\affil{Georgia Institute of Technology, School of Computational Science and Engineering, Atlanta, USA}

\maketitle


\begin{receivedinfo}%
~
\end{receivedinfo}

\begin{abstract}
    In electromagnetism, the model of Maxwell’s equations yields accurate and trustworthy predictions. Numerical solvers can reach electromagnetic solutions far beyond the set of analytical closed-form solutions; this is crucial in metamaterials design where the goal is to find the geometry that generates an optimal electromagnetic solution for a desired property. Then why do we still need to create computational methodologies in metamaterials research? And why should we care about analytical models? The simulation and optimization bottlenecks limit the range of applications of the state of the art. Hence, there is room for opportunities to create new computational methodologies. Semi-analytical methods that are customized for classes of applications enable simulation and optimization. We detail the state of the research in customized methodologies using the illustrative example of optical metasurfaces and suggest best practices. We also discuss how machine learning may enhance these methodologies.
\end{abstract}


\section*{Introduction}

Computing has become a necessary tool for scientific discovery in electromagnetism. Specifically, numerical simulation and optimization enable inverse design—a method that optimizes a desired functionality to find an optimal device geometry automatically~\cite{ molesky2018inverse}. Inverse design has transformed multiple areas of integrated and free-space optics for the direct optimization of physical properties~\cite{ christiansen2021inverse, elsawy2020numerical, li2022empowering} or complex tasks like optical computing~\cite{wetzstein2020inference, hu2024diffractive, li2024exploring}.  Maxwell's equations are at the core of inverse design successes in electromagnetism because they describe the behavior of electromagnetic fields very reliably; discrepancies between full-wave simulations and experimental realizations are attributed to fabrication uncertainties rather than modeling error. Yet, analytical models that simplify the Maxwell's equations continue to guide intuition and innovations.

If simulations can replace experiments and optimization can replace the process of trial and error of the scientific method, then why has computing not ended the use of analytical models in electromagnetism? Analytical and semi-analytical models are still relevant because the one-size-fits-all computational implementation of Maxwell's equations suffers from a simulation bottleneck, despite recent advances and hardware accelerations~\cite{hughes2021perspective}.

\begin{figure}
    \begin{center}
        \includegraphics[width=\textwidth]{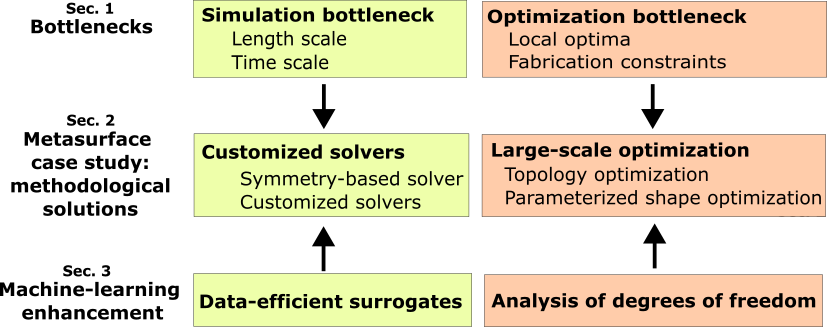}
        \caption{Diagram of the article. Simulation and optimization bottlenecks justify the need to create a customized approximate solver and enable inverse design. These methods may be enhanced by machine learning.}\label{fig:diagram}
    \end{center}
\end{figure}

Here, we focus on the simulation and optimization methods for metamaterials, where the optimizer searches a potentially large-scale geometry that results in a desired property or functionality. Fig.~\ref{fig:diagram} shows an outline of the article. \edit{Although the discussions of approximate solvers and design methods may apply broadly to electromagnetic and acoustic design, the scope of applications in this article are focused on photonics.} In the current state of the art of numerical simulation and optimization, the frontier of applications solved by inverse design is limited by the computational costs of the methods; we call these limits the simulation and optimization bottlenecks (Sec.~\ref{sec:BN}). At the edge of computing capabilities, customized computational methods are mixed with analytical methods and approximations into customized approximate solvers for specific applications, such as metasurfaces. We highlight the main computational approaches to inverse design for optical metasurfaces and suggest standard practices (Sec.~\ref{sec:MS}). We also offer perspectives on where machine learning could enhance and advance numerical methods and under what conditions (Sec.~\ref{sec:ML}). In the concluding remarks (Sec.~\ref{sec:ccl}), we present exciting ways forward and opportunities for future research.

\section{Bottlenecks}\label{sec:BN}

\subsection{Simulation bottleneck}
Large time and length scales are the main bottlenecks to simulating Maxwell's equations. To solve these bottlenecks, semi-analytical methods make the simulation tractable and enable the optimization of a desired property.

Several important optical phenomena happen over a very long time. For example, due to the weak nonlinearities in optics, simulations need to run over long time scales to study or design their effects. For Raman emission and second harmonic generation, semi-analytical methods transform the nonlinear Maxwell simulation with long time scales into two linear Maxwell's simulations in the frequency domain and enable inverse design~\cite{lin2016cavity}. The propagation of solitons in a nonlinear medium can be reduced into a one-dimensional nonlinear Schrödinger equation~\cite{hasegawa2013optical}. In lasing systems, the time scale of the optics is much faster than the time scale of the excited populations in the Maxwell-Bloch equations~\cite{haken1985laser}, resulting in a very long optical time scale. However, the approximation of steady-state ab initio laser theory separates the time scales and enables faster simulation~\cite{ge2010steady}.

Length scale is also a very limiting factor, because the resolution of a simulation is constrained by the wavelength while the size of the computational domain may be very large. In integrated photonics, to account for overall effect on a layout or to simulate a waveguide over long distances with a gradual bend, the semi-analytical method of the coupled wave equation~\cite{marcuse2013theory, katsenelenbaum1998theory, doerr2008dielectric} is needed to accommodate the length scale. Symmetries can also reduce the effective size of the computational domain and enable a larger scale; Refs.~\cite{christiansen2020fullwave} and \cite{sun2025scalable} leverage axi-symmetry and discrete axi-symmetry, respectively, to enable the optimization at the scale of one thousand wavelengths in diameter. In this article, we will detail the approximations that enable the simulation of metamaterials, and particularly metasurfaces (Sec.~\ref{sec:MS}).

\subsection{Optimization bottleneck}
\begin{figure}
    \begin{center}
        \includegraphics[width = \textwidth]{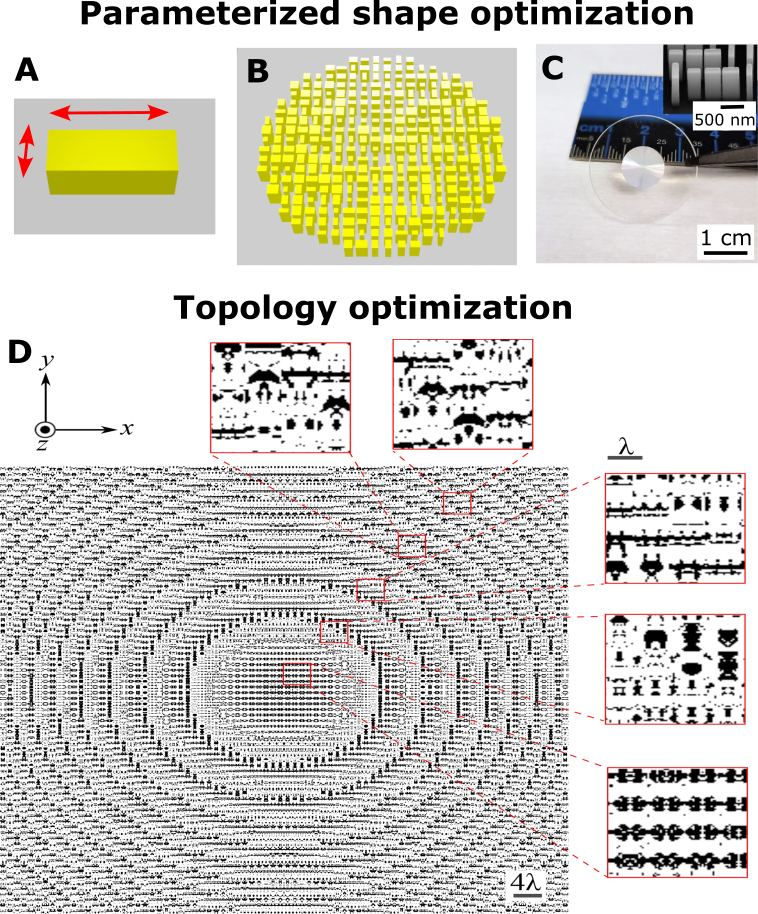}
        \caption{ Parameterization choices. Shape optimization relies on (A) parameterized meta-atom (red arrows depict the parameters) arranged into (B) a metamaterial structure. A and B are adapted from Ref.~\cite{pestourie2020active}. (C) Illustrative metasurface from Ref.~\cite{li2022inverse} that clearly shows a large scale with a 1-cm diameter, and (inset) a small scale with the meta-atom of 500-nm size. (D) Example of topology-optimized structure from Ref.~\cite{lin2019topology} that illustrates the diversity of attainable designs (insets) at the risk of creating an unfabricable design, when fabrication constraints are not enforced explicitly.}\label{fig:MS}
    \end{center}
\end{figure}

The main optimization bottleneck stems from the nonconvexity of most objective functions and the high dimension of the optimization feasible space~\cite{boyd2004convex}. In multiple dimensions, only partial ordering is available, and defining an optimum requires the definition of tradeoffs. We will limit our discussion to the optimization of a scalar function that represents a property of a Maxwell's solution, where an optimum is defined without tradeoffs. For problems with few ($\ll100$) optimizable parameters, inverse design is possible with global optimization techniques which converge slowly. Evolution algorithms~\cite{elsawy2020numerical, li2022empowering} and generative algorithms~\cite{jiang2021deep, kudyshev2020machine} enabled the discovery of design with exotic properties. For many optimizable parameters ($>100$), designs are limited design to local optima~\cite{boyd2004convex}. Metamaterials usually present many independent geometric degrees of freedom. For example, many pillars, arranged in a lattice, parameterize a metasurface (Fig.~\ref{fig:MS}A and B). All the possible combinations of parameters constitute the feasible space~\cite{boyd2004convex}. The optimizer finds an optimal point of the feasible space for a desired objective. Since the desired objectives are often non-convex and the feasible space is high-dimensional, a design that optimizes the objective over all feasible points is unattainable in a reasonable amount of time~\cite{boyd2004convex}. Instead, the optimizer can find designs that are locally optimal in the feasible space, i.e., local optima, which depend on the optimizer's starting points.

Fortunately, when the objectives are differentiable functions, gradient-based optimization converges in a few hundred iterations to a local optimum using the gradient information~\cite{ruszczynski2011nonlinear}. Adding more parameters will not make the optimizer slower. The adjoint method computes the gradient of a scalar function efficiently at the cost of at most one additional simulation~\cite{johnson2007adjoint}. The cost is remarkably independent of the number of optimizable parameters. Empirically with large dimensional space, local optima tend to perform well~\cite{molesky2018inverse}; this finding was studied rigorously in the deep learning literature~\cite{kawaguchi2016deep, dauphin2014identifying}. However, for problems where there exist poor local optima, it is possible to try different optimizer's initialization until a ``performant'' design is found.  How is performance defined? For several applications, theoretical bounds may be derived from simplifying Maxwell's equation down to a convex problem that can be optimized globally~\cite{chao2022physical}. The derived bounds can help evaluate the performance of the local optimum, and often the optimization performance approaches these bounds. Unfortunately, they are not guaranteed to be tight bounds.

\section{Case study of optical metasurfaces}\label{sec:MS}

Metasurfaces are thin metamaterial devices that are patterned at the subwavelength scale to obtain desired physical properties or functionalities. In contrast to thicker optical devices that rely on propagative properties of the materials, metasurfaces rely on the scattering of its geometry features. References~\cite{chen2016review, lalanne2017metalenses, chen2020flat, arbabi2023advances,  kuznetsov2024roadmap} review the many applications of optical metasurfaces from the past decade.
Two length scales of the metasurface geometries create the simulation bottleneck. As the metasurface diameter—the large length scale—increases to up to centimeters (Fig. ~\ref{fig:MS}C), this simulation becomes prohibitive because the resolution is set by the subwavelength features—of the small length scale ($\sim$100s nm)—that are orders of magnitude smaller than the diameter, Fig.~\ref{fig:MS}C (inset). \edit{To illustrate that a metasurface design requires a large-scale optimization, consider the 1-cm diameter in Fig.~\ref{fig:MS}C. The diameter is 25,000-unit-cells long: a total of about half a billion unit cells that results in one billion parameters to optimize a single design.}

Ref.~\cite{yu2011light} presented a methodology that solved this bottleneck by separating simulation from optimization. Instead of simulating, this method starts with a known analytical solution to the problem and then infers ideal conditions that the designer must match locally on the metasurface. This method is called ``phase matching'' because it often approximates the ideal field by its phase (assuming constant amplitude). The approximation of local periodicity makes it possible to build a dictionary that links parameters to the phase, in function of the feature parameters. The dictionary accelerates the search of the geometry features—called meta-atoms here on out—that would realize the needed phases. For example, the parameters may be the length and width of a nanofin (red arrows in Fig.~\ref{fig:MS}A). The optimization then consists of finding the parameters of each meta-atom independently to match the ideal phase function. 

Ref.~\cite{li2022empowering} discussed the drawbacks of this approach: it is limited to known solutions that optimize a desired property (e.g., lensing, beam shaping). More importantly, it breaks down when there are no parameters that map to the needed ideal local phase, which happens as the application of the metasurface becomes more complex (e.g., multifunctional metalenses~\cite{kuznetsov2024roadmap}).

With the same locally periodic approximation, the creation of an approximate solver lifts the fundamental barrier of the phase matching method~\cite{pestourie2020assume}. Ref. ~\cite{pestourie2018inverse} introduced a such customized solver that addresses the bottleneck of scalable metasurface simulations. The creation of a customized solver has enabled the direct optimization of the desired property--—also known as inverse design---and ensures four key benefits. 

First, inverse design ensures pragmaticism: instead of imposing a potentially unattainable ideal solution, the optimizer \emph{discovers the optimal attainable solution and the optimal parameters for the geometry, simultaneously}. Second, the optimization space is the space of geometry parameters rather than that of the electromagnetic fields, i.e., \emph{the feasible space is a fabricable space}. Third, the customized solver exploits synergies between meta-atoms as it \emph{enables the optimization at the large scale of the metasurface}, in contrast to the small scale of the meta-atom. Fourth, the optimizer automatically discovers the fields that are optimal for the desired property without a priori knowledge of the ideal solution. The customized solver \emph{opens the door to complex applications}, such as end-to-end optimization~\cite{lin2022end} for statistical inference~\cite{wetzstein2020inference}.

Incidentally, this paradigm, that shifts from ideal fields to a customized solver, transfers well to radiofrequency (RF) metasurfaces. In RF, the generalized sheet transition condition optimizes an ideal susceptibility that is hard to match with a fabricable device geometry~\cite{achouri2015general}. Customized solvers enable the direct optimization of the device geometry~\cite{budhu2024new}, while discovering the attainable electromagnetic fields that best solve the problem.
\subsection{Benefit of customized solvers}

\begin{table}\label{tab:OOM}
    \centering
    \begin{tabular}{|l|l|}
        \hline
        \textbf{Methods for 1-cm optical metasurface} &
        \textbf{Simulation time} \\ \hline
        GPU-accelerated full-wave solver~\cite{hughes2021perspective} &
        $\sim$ months \\ \hline
        Approximate solver~\cite{lin2019topology} &
        $\sim$ days \\ \hline
        Surrogate-accelerated approximate solver~\cite{li2022inverse} &
        $\sim$ seconds \\ \hline 
    \end{tabular}
    \caption{Rough orders of magnitude in speed for the simulation of an optical metasurface with 1~cm diameter~\cite{li2022inverse}. For Refs.~\cite{hughes2021perspective,lin2019topology}, the speed estimates assume linear scaling in the surface of the metalens.}
\end{table}

Table~1 shows estimates of time comparison between different simulation methods for the 1~cm-diameter metasurface of Fig. ~\ref{fig:MS}C~\cite{li2022inverse}. We assume linear scaling of the computing time with the area of the metasurface for state-of-the-art simulation techniques of a GPU-accelerated finite-difference time-domain solver~\cite{hughes2021perspective} and the customized solver of Ref.~\cite{lin2019topology}. Table~1 illustrates the tradeoff between versatility and speed, and motivates how customized solvers may reduce simulation time by orders of magnitude.

Full-wave solvers are accurate and versatile because they solve Maxwell's equations without any assumptions; however, they would take months to simulate the 1~cm-diameter metasurface once, making the cost of inverse design too prohibitive for this device.

The customized solver from Ref.~\cite{lin2019topology} is based on the locally periodic approximation mentioned above and brings the simulation time to about a day. This customized solver is valid only when the fields are slowly varying along the metasurface. Its acceleration comes from breaking down the large computational domain into small subdomains that are solved independently in an embarrassingly parallel way.
If we further restrict the versatility to when the geometry of the subdomain has few parameters ($\ll100$) and to when few quantities describe the simulated physical response, then the function that maps the geometry parameters to the physical response can be approximated by a data-driven model—a surrogate model. For example, in far-field computations with meta-atoms of subwavelength period, only the zeroth diffractive order propagates. Thus, the transmitted field of the meta-atom can be summarized with a single complex coefficient---the complex transmission coefficient~\cite{pestourie2018inverse}. The surrogate model evaluates $10^4-10^6$ faster than solving for Maxwell's equations for 3D simulations~\cite{pestourie2018inverse, pestourie2020assume, pestourie2020active, pestourie2023physics} and brings the simulation time down to seconds. However, there is no free lunch; the speedup at evaluation time comes at the cost of narrowing down the scope, where the solver is accurate, and training a surrogate model. As the number of parameters increases, the need for training data grows exponentially; this is known as the curse of dimensionality~\cite{bellman2015applied}. In the perspectives section (Sec 3.1), we highlight work about data-efficient surrogate models to increase the number of parameters.

\subsection{The optimization bottleneck of fabrication constraints}

The fabrication of optical metasurfaces is compatible with semiconductor foundry technologies which can scale at a low price. Designs with a growing level of complexity 
are attainable and give the designer room for choices in terms of parameterization of the geometry. \edit{Features at a scale of a few dozens nm can be patterned over areas of diameters up to centimeters. For example, lithography techniques can easily achieve 70~nm resolution; this resolution can be brought down to $\sim40$~nm with process optimization~\cite{song2022study}}. The parameterization of the metamaterials is a choice that impacts the customized solver and potentially the design performance and fabricability. We highlight two classes of parameterizations: parameterized shape optimization and topology optimization.

Parameterized shape optimization (PSO), not to be confused with shape optimization~\cite{sokolowski1992introduction}, breaks down the geometry of the metasurface into meta-atoms of specific parameterizable shapes (squares, rectangles, circles) as illustrated by a nanofin meta-atom in Fig.~\ref{fig:MS}A and its corresponding aperiodically patterned metasurface in Fig.~\ref{fig:MS}B. A central benefit is that the parameterized shapes can be exploited by a customized approximate solver that simulate each meta-atom separately. PSO also builds the fabrication constraints such as the minimum feature size into the parameterization. For example, for a minimum feature size of 100~nm and a nanofin parameterization (Fig.~\ref{fig:MS}A), the lower bound of the parameters is set to 100~nm and the upper bound is set to $(\Lambda -100)$~nm, where $\Lambda$ is the period of the meta-atom. However, PSO constrains the types of designs to the predefined patterns. For example, in the nanofin parameterization (Fig.~\ref{fig:MS}B), no pillar can extend to multiple meta-atoms. Therefore, PSO limits the designs to a subset of fabricable geometries and may leave out more performant designs. \edit{To be precise, the choice of basis for the parameterization of a design function has no effect if the basis is complete. However, it may impact performance when the parameterization also involves a reduction in the number of degrees of freedom. Measuring the effect a lower-dimensional parameterization is challenging. Very low dimensions ($< 10$) enable extensive global searches; the number of parameters can be iteratively increased using successive refinement \cite{ mutapcic2009robust, pestourie2018inverse}, but this approach does not ensure better performance than direct optimization in a feasible space of larger dimension with random initial guesses. Parameterization in shape optimization is still an active research area~\cite{ zhou2024large, dainese2024shape}; we also share a perspective on degrees of freedom in ML-enhanced optimization in Sec.~3.2.} \edit{The wavelength of design affects the choice of solver and fabrication methods. For design with longer wavelength, devices are smaller compared to the wavelength because they tend to not scale proportionally. Scalable fabrication processes like 3D printing become available~\cite{roques2022toward}, and full-wave simulation and topology optimization are more tractable. For shorter wavelength, such as extreme UV or X-ray, it is challenging to fabricate feature that are comparable to the wavelength, in that case local ray optics (e.g., X-ray coded apertures~\cite{caroli1987coded}) or diffractive optics~\cite{o2004diffractive} may prevail, however LPA including diffractive orders~\cite{pestourie2018inverse} may present a tradeoff that enable shape optimization.}
More degrees of freedom in the geometry parameterization may result in increased performance; this is called the blessing of dimensionality~\cite{gershenfeld1999nature}.  

To increase the degrees of freedom, freeform topology optimization (TO) starts from the optimization of a density of pixels to which the designer allocates a material. TO can generate arbitrary shapes~\cite{jensen2011topology, christiansen2021inverse, molesky2018inverse}, but fabrication constraints are no longer built into the parameterization. Fig.~\ref{fig:MS}D illustrates an example of design resulting from a density-based topology optimization that does not enforce fabrication constraints~\cite{lin2019topology}. Notice the size of the pixel is much smaller than the fabrication minimum feature size of 50~nm, yet Fig.~\ref{fig:MS}D insets show features that are as small as a single pixel. Fig.~\ref{fig:MS}D clearly illustrates that the feasible space of TO is no longer a fabricable space. So, fabrication constraints must be imposed during the optimization via filtering, thresholding, and nonlinear constraints~\cite{hammond2021photonic, chen2024validation, christiansen2021inverse}. Although the state of the art of topology optimization already integrates fabrication constraints~\cite{hammond2021photonic, chen2024validation, christiansen2021inverse}, the hyperparameters of their implementations (constraints' coefficients and how quickly to impose them) may significantly impact the overall design performance. TO performs gradient-based optimization on artificial materials that are interpolated between the materials that the designer needs to choose. The binarization constraint---that forces the choice of one material instead of the other---is gradually turned on~\cite{christiansen2021inverse}, and interpolation between metals and dielectrics requires special care~\cite{christiansen2019non}.

A customized solver may bring TO to large diameters at the cost of approximation errors~\cite{lin2019topology}. However, the density parameterization of TO needs many inputs variables. Therefore, the potential costs savings of a surrogate model acceleration may be overshadowed by the data generation costs and the curse of dimensionality~\cite{bellman2015applied}.
Quantifying how much performance would improve with the transition from shape optimization to topology optimization is an open question that is certainly application dependent. However, the fundamental role of thickness has been elucidated~\cite{miller2023optics} and empirically illustrated with topology-optimized designs that gain performance from larger volumes~\cite{lin2021computational}.

Incidentally, TO can be used for shape optimization when the shape parameterization is differentiable with respect to the refractive index~\cite{mansouree2021large}. However, this approach suffers from the same scaling limits as full-wave simulations.
The choice of parameterization constrains the choice of approximation to build the customized solver. We now review the approximations and their corresponding customized solvers in the context of metasurfaces.

\subsection{Customized solvers break the simulation bottleneck}

\begin{table}\label{tab:ap}
    \centering
    \begin{tabular}{|p{0.25\linewidth}|p{0.22\linewidth}|p{.22\linewidth}|p{0.18\linewidth}|}
    \hline
    \textbf{Approximation method} &
    Periodic approximation (LPA~\cite{pestourie2018inverse}, nonlocal~\cite{overvig2024spatio}) &
    Absorbing~\cite{phan2019high} and overlapping domains~\cite{lin2019overlapping} &
    Axi-symmetric full-wave solver~\cite{christiansen2020fullwave, sun2025scalable}\\ \hline 
    \textbf{Surrogate acceleration} &
    Yes & 
    No &
    No \\ \hline
    \textbf{Strengths} &
    Fastest and reliable for nearly periodic fields.&
    Embarrassingly parallel. Accurate local field effects. &
    Very accurate (no approximation).  \\ \hline
    \textbf{Limits} &
    Breaks down for high NA and small glancing angles. &
    Cannot compute long-lived resonances. &
    Scales the least. (Discrete) Axi-symmetric design. \\ \hline
    \end{tabular}
    \caption{Summary of approximate customized solvers for optical metasurfaces.}
\end{table}

Physical approximations help make the simulation of metasurfaces tractable as they break the large simulation---with a diameter of thousands of wavelengths, as illustrated in Fig.~\ref{fig:MS}C---into simulations with a significantly smaller diameter (illustrated in Fig.~\ref{fig:MS}A). Different approximations create different local fields at the level of the metasurface. In all cases, the local field generates a solution in the far field using a near-to-far-field transformation. \edit{Where the local approximation is propagated via a convolution with the appropriate Green's function~\cite{pestourie2018inverse}
\begin{equation}
    e(x)=\int_V G(x, x^\prime)f(x^\prime)dx^\prime
\end{equation}
where, in its most general form, $G(x, x^\prime)$ is a $6\times 6$ tensor, and $e$ is the electromagnetic field (6 components), and $f$ represents the electric and magnetic currents (6 components) that are equivalent to the local field. Note that if the support of the integral $V$ is an infinite plane and the propagated field is only needed in the upper half space above the source plane (e.g., in transmission or reflection only), subtracting the mirror symmetric fields does not change the field above, however, it cancels the electric current source and reduces the needed components of the Green's function down to $6 \times 3$~\cite{pestourie2018inverse}.}

If there was a periodic symmetry in the geometry of the metasurface,
the whole metasurface would be accurately simulated with a computational domain of the size of a single meta-atom (illustrated in Fig.~\ref{fig:MS}A) with periodic boundary conditions. This approach has a long history for lensing applications~\cite{lalanne2017metalenses}. For aperiodic metasurfaces, computing the field for each meta-atoms with periodic boundary conditions is equivalent to assuming that the rate of change in geometries from a meta-atom to its neighbor is negligible to the zeroth order. 
For metasurfaces, this approximation is called the locally periodic approximation (LPA)~\cite{pestourie2018inverse} (Table~2 Column 1), where the field computed for each meta-atom is a local approximation of the transmitted (or reflected) field by the metasurface. This approximation of the local field breaks when the scattered field changes significantly from one meta-atom to the next, such as high-numerical-aperture applications (high-NA)~\cite{chung2020high}, or at small glancing angle~\cite{perez2018sideways}. Incidentally, when building a surrogate model for the subdomain simulations, it is beneficial to smear out the resonances that come from the periodic simulations as in Ref.~\cite[SI]{li2022inverse}, because the effective lack of periodicity will likely attenuate the quality factor Q of the actual device. \edit{The Q factor impacts the accuracy of LPA by reducing the tolerable rate of change. As a rule of thumb, a design from an approximate solver that relies on a high Q factor, a high non-local range, and a high dispersion, is more vulnerable to the rate of change in the designed scattered field. Note that the rate of change in the scattered field matters more than the rate of change in geometry. In the case of guided-mode resonances, Ref.~\cite{ fisher2022efficient} explicitly adds coupled mode theory on top of LPA to account for nonlocal effects. The case of leaky resonances is treated in Ref.~\cite{ huang2023leaky}.} \edit{The index contrast of a chosen parameterization does not affect the accuracy of LPA directly. Higher index contrasts include feasible designs that could break the approximation. However, the optimality of these designs, especially, whether the problematic designs are chosen or not, is dictated by the application. A lens application requires slowly varying fields that are typically suited for LPA, whereas applications like angle detection or spectroscopy might result in designs that break the approximation. Low index contrasts, illustrated in Refs.~\cite{nikkhah2024inverse, roberts20233d}, may increase the overlap between feasible designs and designs that are valid for LPA.}
In contrast, nonlocal metasurfaces take advantage of the resonant modes to enable additional functionalities~\cite{overvig2020multifunctional,
 shastri2023nonlocal}. They also assume small perturbations in the rate of change, so that the resonant mode is not affected (zeroth order in the rate of change). \edit{The engineered perturbations produce designs that preserve optical nonlocality \cite{zhou2023multiresonant}.} For nonlocal metasurfaces, the approximation breaks down when the perturbation is big enough to significantly change the resonant mode.
Recently, Ref.~\cite{overvig2024spatio} unified the theory for local and nonlocal metasurfaces and introduced a framework that allows for surrogate models in both cases.
Below, we highlight an emerging method that is first-order accurate in the rate of change of the metasurface and enables the computation of the coupling between waveguide modes and the free space (in the zeroth order approximation, guided modes do not couple to free space).

Instead of breaking the large metasurface simulation into subdomains with periodic boundary conditions, other works broke the domain into subdomains with absorbing boundary conditions (Table~2 Column 2) as in Ref.~\cite{phan2019high}. This approximate solver benefits from using a subdomain with diameters larger than the wavelength. The distributed T-matrix simulation method~\cite{skarda2022low} is also like an absorbing boundary condition by neglecting long range scattering. It was shown to be more accurate than the LPA-customized solver on metasurfaces with meta-atoms of high aspect-ratio scatterers of a large period~\cite{skarda2022low}.
Including a small portion of the neighboring subdomains ($\sim \frac{\lambda}{10}$ of the neighbor) gains additional accuracy and ensures better estimation of the field at the edge of the subdomains. This is called the overlapping domain approximation~\cite{lin2019overlapping}. These methods can produce more accurate simulations and better design for high-NA lenses~\cite{phan2019high, lin2019overlapping}. However, these approximations will break down with long-lived modes that are larger than the subdomains. These methods are more accurate when the subdomain is large compared to the wavelength. This implies that there are more parameters to describe the geometry on each subdomain where the accelerations via surrogate models (Table~1) suffers from the curse of dimensionality~\cite{bellman2015applied}.
Note that in the context of ``phase matching,'' similar methods to overlapping domains have included the neighboring features in the local simulations to improve the accuracy of the phase approximation~\cite{hsu2017local, an2022deep}.

We summarize this discussion about the state of the art of approximate solvers in Table~2.

\subsection{Suggested practices}

\begin{table}\label{tab:sug}
    \centering
    \begin{tabular}{|p{0.3\linewidth}|p{0.2\linewidth}|p{.2\linewidth}|p{0.17\linewidth}|}
    \hline
    \textbf{Device/Application} & \textbf{Suggested method} &
    \textbf{Computing resources} &
    \textbf{Maximum diameter} \\ \hline
    Metasurfaces $\sim$ normal incidence: beam forming/ steering, lensing. &
    LPA/ nonlocal surrogate-based solver &
    Minutes on laptop &
    $>10000 \lambda s$ \\ \hline
    Metasurfaces with high NA or blazing angle. &
    Overlapping/absorbing domains &
    $\sim 40$ computing units for hours &
    $\sim 1000 \lambda$s \\ \hline
    Metasurface with large optical volume (multiple layers coupled in near field) &
    Full-wave/ axi-symmetric solver (no approximation) &
    $>40$ computing units for hours to days &
    $\sim 100 \lambda$s ($1000 \lambda$s if axi-symmetric) \\ \hline
    Grating coupler over small volume/ integrated photonics &
    Full-wave solver &
    $\sim 40$ computing units for hours to days &
    $\sim 100 \lambda$s \\ \hline
    \end{tabular}
    \caption{Suggested computational methods for selected applications}
\end{table}

In Table~3, we suggest simulation methods for selected applications. For thin optical metasurfaces, where the desired field is ``nearly'' periodic, and the incident angle is near normal incidence, we suggest using surrogate-based solvers based on LPA. They are orders of magnitude faster than full-wave simulations and their approximation seems accurate enough in these settings. Numerous designs of beam-steering or beam-forming, as well as lensing, have used this approximation (either via surrogate-based solvers or ``phase matching'') and were experimentally validated~\cite{kuznetsov2024roadmap}. The design space can scale to diameters of 10,000s of wavelengths, and an optimization run takes minutes on a laptop~\cite{li2022inverse}.

LPA breaks down for high-NA~\cite{chung2020high} or small glancing angle~\cite{perez2018sideways}. For these metasurface applications, overlapping domains~\cite{lin2019topology} and absorbing domains~\cite{phan2019high} have shown highly performing designs that were validated numerically. These methods perform better for large subdomains. Their many geometrical parameters render surrogate models impractical. However, these methods are embarrassingly parallel. The design space may scale up to 1,000 wavelengths and a design run may take hours on a few dozen computing units.

Large optical volumes help increase performance~\cite{miller2023optics}. With larger optical volume, the optical path length increases inside the design domain and it remains an open question whether customized solvers can become accurate in these settings. We suggest using full-wave solvers, which may be GPU accelerated though not embarrassingly parallel~\cite{hughes2021perspective}. Diameters of 100s of wavelengths and up to 1,000 wavelengths with axi-symmetry~\cite{christiansen2020fullwave, xue2023fullwave} or discrete axi-symmetry~\cite{sun2025scalable} can be designed in hours to days using a few dozen computing units. 

To finish, although it is not a metasurface, we suggest full-wave simulations for integrated photonics~\cite{molesky2018inverse}. When the goal is to miniaturize a component like a grating coupler, the diameter length scale is less problematic. We suggest a design domain size of up to 100 wavelengths using a full-wave solver. The design run may take hours to days on a few dozen computing units.

\subsection{Emerging customized methods}
We now highlight recent work on methodologies that may enable new applications for incoherent light and coupling free space to guided modes.

Metasurfaces are often designed for coherent light, where a phase is well defined. However, major practical light sources are incoherent (e.g., black body radiation such as the sunlight). The incoherent light is a random light that follows a statistical distribution~\cite{wolf2007introduction}. For example, it can be described as an expectation $\int d\theta f(\theta) E(\theta) $ where $E(\theta)$ is the field evaluation from Maxwell's equations at a given angle. In common cases, $1000$ quadrature points are needed to compute this average accurately~\cite{pestourie2022efficient} thus multiplying the cost of one simulation by as much. However, in practice since each simulation is computing-resource intensive, only a few simulations are performed to estimate the average~\cite{scranton2014single} at the cost of accuracy. Fortunately, when the physical property to be averaged can be cast as a linear function of the field that has rank $k$, only $k$ reciprocal simulations are needed to compute the average effect without loss of accuracy and the integral can be computed efficiently (sometimes analytically)~\cite{yao2022trace}. This approach was showcased for the efficient design of metasurface collimators and concentrators of incoherent light with a thousand-fold speedup that is independent of the solver chosen~\cite{pestourie2022efficient}.

Coupling light from free space to guided optics is a key application in nanophotonics~\cite{cheng2020grating}. However, as the size of the device increases, the simulation becomes a bottleneck of the inverse design. For the specific parameterization of a patterned waveguide designed to enhance the coupling between free space and guided modes (a metasurface on top of a waveguide), a customized surrogate-based solver may accelerate simulations. Unfortunately, the LPA cannot account for the coupling because guided modes do not couple light to free space. Normalizing free-space modes is also a challenge~\cite{fisher2022efficient}. Ref.~\cite{fisher2022efficient} solves this issue by combining the LPA, the coupled mode theory, and the perturbation theory to lay the foundations of a fast solver that is accurate to first order with the rate of change of the metasurface patterns. \edit{Instead of guided modes, leaky resonances may also be used to merge free-space optics and integrated optics~\cite{huang2023leaky}.}

To conclude this section, customized solvers provide significant scaling and savings. However, their approximations also limit their range of applications. Extending the reach of customized solvers is an open question that may significantly impact the field of inverse design by enabling designs at scale.

\section{Machine-learning enhancements}\label{sec:ML}

\begin{figure}
    \begin{center}
        \includegraphics[width=\linewidth]{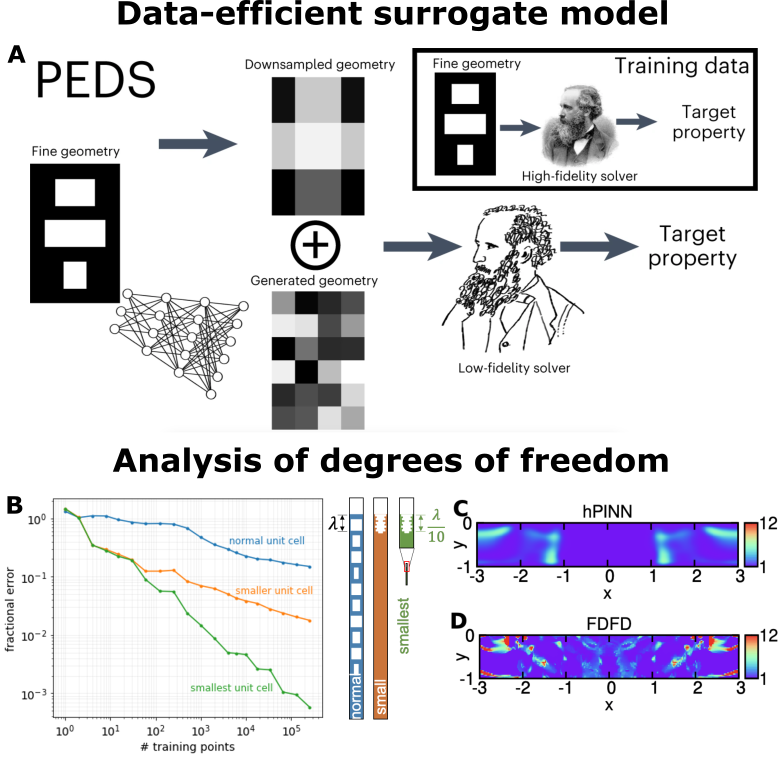}
    \end{center}
    \caption{Machine learning can enhance inverse design in electromagnetism by accelerating the customized solver and analyzing the degrees of freedom of design problems. (A) A physics-enhanced deep surrogate improves the data efficiency of surrogate models by leveraging a Maxwell’s solver layer~\cite{pestourie2023physics}. (B) Training curves of a surrogate model illustrates that unit-cell have different numbers of physical degrees of freedom even though they have the same number of geometrical degrees of freedom~\cite{pestourie2020assume}. (C and D) Designs from a neural network (hpinn) and from topology optimization (FDFD) look very different despite having the same performance: ML-enhanced design may present lower-frequency special features~\cite{lu2021physics}.}\label{fig:ML}
\end{figure}

\subsection{Surrogate models }
Table~1 shows that the main benefit of data-driven models is to speed up the customized approximate solvers by orders of magnitude. Surrogate models are fast because they evaluate a fit, while solvers need to solve Maxwell's equations. However, their fast evaluation comes at the cost of training. Let $T$ be the training cost; the training costs is often dominated by the data generation costs which requires to simulate Maxwell's simulations for many points in the feasible space. $E_{solver}$ and $E_{surrogate}$ are the evaluation costs of computing the property from a solver solution or the evaluation of a trained surrogate, respectively. $E_{surrogate}$ is $10^3\times$ to $10^6\times$ cheaper than $E_{solver}$~\cite{pestourie2018inverse, pestourie2020assume, pestourie2020active, pestourie2023physics}. Reusability $R$ must amortize the training cost to justify the usage of a data-driven surrogate, such that

\begin{equation}
R \times E_{surrogate} + T \ll R \times E_{solver}.
\end{equation} 

For customized approximate solvers, $R\sim L N$, where $N$ is the number of meta-atoms from hundreds to millions~\cite{pestourie2018inverse, bayati2020inverse, bayati2022inverse, li2022inverse}, and $L$ is the number of optimization iterations needed to converge to a local optimum ($\sim 100$--$1000$), it makes perfect use cases for data-driven surrogate models. Using data-driven models to accelerate the simulations in large-scale optimization is aligned with the benefit of machine learning for accelerating simulations, highlighted in Ref.~\cite{woldseth2022use}.

For up to 5 parameters, Chebyshev interpolation is an excellent data-driven surrogate model~\cite{pestourie2018inverse, bayati2020inverse, bayati2022inverse, li2022inverse}  because of its exponential convergence for smooth functions~\cite{boyd2001chebyshev}. However, when the number of parameters of the surrogate model increases, so does the number of data needed, exponentially. Neural networks may mitigate this curse of dimensionality~\cite{cheridito2021efficient}. Neural surrogate models are not to be confused with neural solvers to partial differential equations~\cite{karniadakis2021physics}. Neural surrogate model only predicts a (low-dimensional) property of the solution as a function of parameters in the feasible space.

Deep learning has been extensively studied in the context of metamaterials~\cite{khatib2021deep, ji2023recent} and photonic design~\cite{ma2021deep, wiecha2021deep}. In the context of surrogate models, Ref.~\cite{pestourie2020active} introduces a Bayesian active learning algorithm that saves one order of magnitude of data by incrementally building a dataset customized for the surrogate model. Purely neural implementations of the surrogate models have the benefit of being easy to implement using software libraries~\cite{abadi2016tensorflow, paszke2019pytorch, pal2022lux}, and they apply to any partial differential equations and target properties. However, they still suffer the curse of dimensionality. More importantly, they relearn the already-known Maxwell’s equations from data, which is a wasteful process.

Inductive bias builds the knowledge of Maxwell’s equation inside the model architecture~\cite{khaireh2023newcomer, liu2024kan}. A physics-enhanced deep surrogates (PEDS) uses a solver for Maxwell's equation as its last layer~\cite{pestourie2023physics}. A solver layer enforces that the output of the model obeys the governing equations and conservation laws and increases interpretability, as illustrated in Fig.~\ref{fig:ML}A. The input is the same parameterization of the geometry as the one used to generate the high-fidelity dataset (Fig.~\ref{fig:ML}A (inset)). A neural network transforms the PEDS input into an input to the solver layer. At that stage, additional field knowledge, i.e., inductive bias (such as symmetry or approximate geometry), can be mixed into the solver’s generated input. Then, the solver performs the simulation and returns the surrogate model prediction. With the Maxwell’s solver layer, PEDS reduces the data need by at least two orders of magnitude to reach few-percent error. In addition, it evaluates $10^2$ to $10^4$ faster than the high-fidelity solver. Why? Because the solver layer is a fast low-fidelity solver (depicted by a drawing of James Maxwell in Fig.~\ref{fig:ML}A). The low-fidelity solver has an unacceptable level of error, but the neural network is trained end-to-end to make the solver layer accurate for the desired property. The low-fidelity solver computes a Maxwell’s solution that is different from the high-fidelity solution, however, it results in the same desired property. An arbitrarily accurate PEDS exists, if the range of desired properties attainable by the high-fidelity model is included in the range attainable by the solver layer.

To conclude, surrogate models accelerate the customized solvers, but they suffer the curse of dimensionality. To improve data efficiency, scientific machine learning includes inductive bias in the model architecture. Data efficiency increases the reach of fast surrogate-based customized solvers. With nine parameters, a customized solver could reach periods bigger than the wavelength that would propagate non-zero diffractive orders. Alternatively, nine parameters suffice for a surrogate-based overlapping domains solver, as in~\cite{lin2019overlapping}, with one parameter per meta-atom and considering a meta-atom and its eight neighbors.

\subsection{Analysis of degrees of freedom}
The ability to reduce the dimensionality of a scalar function may improve not only its learning, but also its optimization. For quadratic functions, the singular value decomposition~\cite{strang2019linear} of their Hessian helps estimate the number of latent dimensions~\cite{pestourie2020active}. There also exist methodologies for nonconvex functions~\cite{van2009dimensionality}. Here, we highlight how neural networks can inform and exploit low dimensionality in nonconvex functions. In this section, we define the geometrical degrees $N_{geom}$ as the dimension of the chosen parameterization of the metamaterials, physical degrees of freedom $N_{phys}$ as the dimension of the latent space that explains most of the nonconvex objective function variations, and the design degrees of freedom $N_{design}$ as the dimension of the latent space that generates performant designs.
 
Unfortunately, the geometrical degrees of freedom don't always match the more important physical degrees of freedom. When $N_{phys}>N_{geom}$, the geometry is underparameterized, whereas when $N_{phys}<N_{geom}$, the geometry is overparameterized. In Fig.~\ref{fig:ML}B, we show the training accuracies of neural-network models with ten geometrical parameters from Ref.~\cite{pestourie2020active}. Each plot line corresponds to a different scale of geometry. The green line, corresponding to a geometry that is 100 times smaller than that of the blue line, reaches an accuracy that is 100 times better for the same number of training points and the same number of geometrical degrees of freedom ($N_{geom}=10$). Two principal components dominate the Hessian of the model for the green curves ($N_{phys}=2$). In that case, the ten geometrical degrees of freedom overparameterize the two physical degrees of freedom.  Physical degrees of freedom help explain how some learning problems are easier than others. For example, in Ref.~\cite{an2020deep}, a scalar function with a $64\times64$ geometry input is learnt with only 100,000 data points. However, the data efficiency may be explained by the fact that the meta-atom has a subwavelength period, resulting in $N_{phys}<<64\times64$. The overparameterization explains the data efficiency.

For a scalar function that represents the property of a PDE solution, multiple input geometries may result in the same outcome. This is known as the many-to-one forward problem and gives rise to the ill-posed many-to-one inverse problem. Thinking in terms of degrees of freedom, the one-to-many forward problem may be beneficial, because it means that there are fewer physical degrees of freedom than geometrical degrees of freedom. The design degrees of freedom are a subset of the physical degrees of freedom that lead to high performing designs: $N_{design}<N_{physics}$. From the many-to-one choices, the designer only needs one of the solutions. Neural networks may be good candidates to learn the ``smoothest'' solution because they learn low-frequency features first~\cite{rahaman2019spectral, herrmann2024neural}. Ref.~\cite{lu2021physics} introduces a method that learns a design representation at the same time as it learns a solver. Fig.~ \ref{fig:ML}C and D show two designs that perform equally well. The design coming from the neural design representation (Fig.~ \ref{fig:ML}C) has a bigger minimum feature size than the design from topology optimization (Fig.~ \ref{fig:ML}D), which may be more conducive to fabrication constraints. The small  $N_{design}$ may also inform a choice of parameterization~\cite{zandehshahvar2022manifold}.

Incidentally, the neural representation that serves as designer and solver in~\cite{lu2021physics} is a physics-informed neural network (PINN)~\cite{raissi2019physics}. In the spirit of Ref.~\cite{mcgreivy2024weak}, we did not list PINN as one of the suggested solvers because they are slow compared to the state of the art of full-wave solvers~\cite{hughes2021perspective}. Nonetheless, PINN may be competitive for inverse problems~\cite{chen2020physics}, not to be confused with inverse desing---the topic of this article.

Versatile and data-efficient ways to discover and learn the design degrees of freedom of nonconvex scalar functions still need to be found. Such method could help analyze how difficult a design goal is (the less design degrees of freedom, the easier). It may also help the designer in choosing the proper geometry parameterization (PSO versus TO) and corresponding customized solver.

\section{Concluding remarks}\label{sec:ccl}

Customized solvers overcome the simulation and optimization bottlenecks using approximations or symmetry to accelerate simulations. These methods enable inverse design, where the optimal geometry and an optimal attainable solution to Maxwell’s equations are co-discovered. Customized solvers represent a tradeoff between specialization and computing-resource efficiency. 

As surrogate models become data-efficient and enable more input parameters, the accurate overlapping-domain approximation may be within reach of surrogate-accelerated customized solvers. Other domain-decomposition approximations may be derived to higher order of accuracy, akin to the Schwarz algorithm~\cite{gander2019class}, to provide additional tradeoffs between speed and accuracy. It remains to be seen whether the LPA is sufficient for novel metasurface applications, such as performing optical inference~\cite{huang2024photonic, wetzstein2020inference}; the performance gain from using a more accurate solver than LPA needs to be studied for end-to-end co-design of sensor and software~\cite{lin2021end}.

\edit{This article focused on methods for photonics, however, approximate} customized solvers are used in frequency regimes away from optical wavelengths. It may continue to enable inverse design in radio-frequency~\cite{budhu2024new} or acoustics~\cite{pestourie2024towards}. Beyond electromagnetism, customized approximate solvers may impact mechanical metamaterials~\cite{bertoldi2017flexible}.

Machine learning may help elucidate the complex question of degrees of freedom in the input space of nonconvex scalar functions and their effects on design performance. The degrees of freedom may shed light on how hard a design problem is. The analysis of design degrees of freedom may inform the choice of parameterization or lead to an efficient design parameterization. Quantifying how the accuracy of a solver impacts the number of physical degrees for a target function is an open question that is central to efficient inverse design with approximate solvers.

\section*{Acknowledgements}
RP thanks Steven G. Johnson for insightful discussions on semi-analytical methods in electromagnetism. RP thanks Georgia Tech for supporting this effort. 


\bibliographystyle{IEEEtran}
\bibliography{refs}



\end{document}